\documentclass[conference]{IEEEtran}
\ifCLASSINFOpdf
\else
\fi

\usepackage{import}
\usepackage{tabularx}
\usepackage{tabulary}
\usepackage{booktabs}

\usepackage{graphicx}
\graphicspath{ {./images/} }

\usepackage{url}

\usepackage{xcolor}
\usepackage{ifthen}
\newboolean{showcomments}
\setboolean{showcomments}{false} 
\ifthenelse{\boolean{showcomments}}
    {
        \newcommand{\song}[1]{\textcolor{gray}{{\it [Song says: #1]}}}
        \newcommand{\emelie}[1]{\textcolor{red}{{\it [Emelie says: #1]}}}
        \newcommand{\per}[1]{\textcolor{blue}{{\it [Per says: #1]}}}

    }
    {
        \newcommand{\song}[1]{}
        \newcommand{\emelie}[1]{}
        \newcommand{\per}[1]{}
    }

\begin{document}
%
\title{Concepts in Testing of Autonomous Systems: \\Academic Literature and Industry Practice}

\author{\IEEEauthorblockN{Qunying Song}
\IEEEauthorblockA{Lund University\\
Lund, Sweden\\
Email: qunying.song@cs.lth.se}
\and
\IEEEauthorblockN{Emelie Engström}
\IEEEauthorblockA{Lund University\\
Lund, Sweden\\
Email: emelie.engstrom@cs.lth.se}
\and
\IEEEauthorblockN{Per Runeson}
\IEEEauthorblockA{Lund University\\
Lund, Sweden\\
Email: per.runeson@cs.lth.se}}


%


\maketitle

\begin{abstract}
Testing of autonomous systems is extremely important as many of them are both safety-critical and security-critical. The architecture and mechanism of such systems are fundamentally different from traditional control software, which appears to operate in more structured environments and are explicitly instructed according to the system design and implementation. To gain a better understanding of autonomous systems practice and facilitate research on testing of such systems, we conducted an exploratory study by synthesizing academic literature with a focus group discussion and interviews with industry practitioners. Based on thematic analysis of the data, we provide a conceptualization of autonomous systems, classifications of challenges and current practices as well as of available techniques and approaches for testing of autonomous systems. Our findings also indicate that more research efforts are required for testing of autonomous systems to improve both the quality and safety aspects of such systems. 
\end{abstract}


%
\IEEEpeerreviewmaketitle

\section{Introduction}
Autonomous systems are expected to replace humans in carrying out a variety of functions \cite{harel2020autonomics}, and can be central and crucial for different industry domains such as automotive, robotics, and aviation. Advances in machine learning and artificial intelligence have enabled an overwhelming progress for such systems. There are already prototypes of autonomous vehicles that are tested on the road, and autonomous systems have replaced humans to a significant extent for decision-making in investment markets, particularly for asset management~\cite{sifakis2019can}.

While autonomous systems are becoming prevalent and have enormous potential for the society, how to test these systems is not resolved yet due to the unpredicted environment they operate in, and their adaptive behaviour \cite{helle2016testing}. One problem is that no industrial standards or common approaches have been settled for testing of such systems \cite{knauss2017paving, agaram2016validation}.  Further, research conducted on autonomous systems tend to be conducted in isolation from industry practice, for example, purely in simulation environments~\cite{chen2018auto}.


To better understand the essence of autonomous systems and the current status of testing of such systems, we conducted an exploratory study by synthesizing academic literature, focus group discussions and interviews with industry practitioners. 
\emph{Our contribution is a synthesis of autonomous systems concepts, their characteristics and functionalities, empirically grounded in research and practice. We also classify challenges, approaches, techniques, and practices available for testing of autonomous systems.} The results indicate that the current state of testing such systems is far from being desirable and it is in need of major improvements. Our synthesis aims at providing tools for industry and academia to align communication on the topic, and to jointly meet the need for more knowledge. 

While similar studies have been conducted in related areas they are either not focusing specifically on autonomous systems, nor on their 
testing. Most of the existing studies we found were either focusing on only one autonomous domain, for example, self-driving cars \cite{knauss2017paving}, or a particular aspect of the system, like safety \cite{borg2018safely}. Our results are comprehensive with respect to the testing of autonomous systems in general, and are inclusive towards both academic research findings and industrial practices. We believe they can serve as a good framework both for future research and industrial development.



 



\section{Related Work}
\label{sec:related}

Helle et al. present an overview of autonomous systems and testing approaches for such systems, mostly from an avionic perspective~\cite{helle2016testing}. In their paper, the authors introduced the concepts and characteristics of autonomous systems as well as the challenges for testing them. They conclude that, due to the dynamically changing environment and system behaviour, conventional testing approaches that aim for fault avoidance, removal, and tolerance are infeasible to ensure the quality of autonomous systems. In addition, they surveyed existing approaches and presented mainly model-based testing approaches and related tools. 
We extend their insights by including also industry practices and by exploring numerous techniques, approaches, and engineering practices for testing of such systems beyond the model-based approaches.

Knauss et al. conducted an empirical study aiming to collect testing challenges for autonomous vehicles~\cite{knauss2017paving}. Similar to our study, they combined a literature review with focus groups, and interviews with both researchers and practitioners. 
Our study is broader, 
focusing on autonomous systems in general and have an explicit goal to extract existing techniques and approaches in addition to the challenges. 

Borg et al.~\cite{borg2018safely} 
present challenges and approaches for testing of deep learning based automotive applications. Their results point to safety cages as a promising solution to be investigated further. The study contains a systematic literature review of 64 papers on safety analysis or verification and validation of machine learning based autonomous cyber-physical systems. Six workshops were conducted with practitioners from the automotive domain, in order to bridge the gap in understanding the state-of-the-art and obstacles on the way forward. 

 Zhang et al. report a systematic literature review on testing and verification of neural-network-based safety-critical cyber-physical systems ~\cite{zhang2020testing}. Their study includes 83 papers from 2011 to 2019. The authors present an overview of different neural networks and a summary of existing approaches for verification of such systems as well as their pros and cons. Another similar study was conducted by Zhang et al. \cite{zhang2020machine} with focus on testing of machine learning systems. In this study, the authors surveyed 144 papers between 2014 and 2019 on testing and verification of machine learning systems, and provided an overview and classifications of techniques and approaches that are employed in research. As a comparison, our study focuses on autonomous systems, independently if they are driven by machine learning technologies or not.

\section{Research Methods}
\label{sec:method}

\begin{figure}[!t]
\centering
\includegraphics[trim=0 8mm 0 8mm, clip, width=\columnwidth]{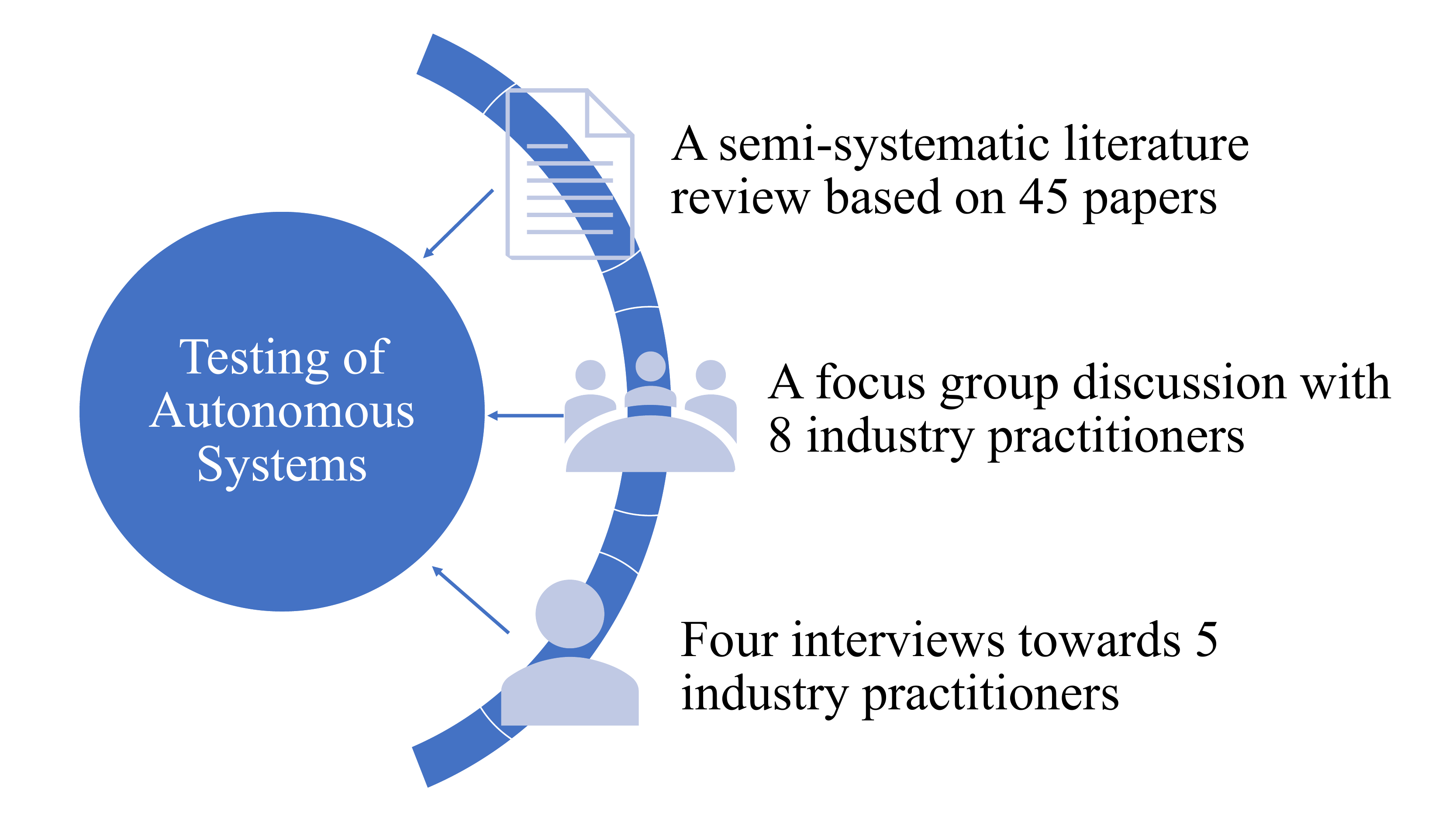}
\caption{Overview of Research Methods}
\label{fig_methods}
\end{figure}

We launched a multi-method study, consisting of 
 a semi-systematic literature review, a focus group discussion, and four interviews, as shown in Figure~\ref{fig_methods}, conducted in the given order. We analyzed the collected data qualitatively and established one thematic model per activity~\cite{runeson2009guidelines}. Then the outcomes from all the three activities were compared and aggregated into a coherent outcome by the end. Our study is guided by three research questions, aimed to build on and complement related work, as defined in Section \ref{sec:related}:

\begin{itemize}
    \item[RQ1] How is the concept of autonomous systems defined?
    \item[RQ2] Which are the principal challenges related to testing of autonomous systems?
    \item[RQ3] What approaches and practices for testing of autonomous systems are used or proposed?
\end{itemize}

\subsection{Literature Review}

\subsubsection{Paper Selection}

We conducted a semi-systematic literature review \cite{snyder2019literature}, where the research questions are broad and the searching and selection process is 
flexible. 
Due to the relative immaturity of the research field, the purpose was to retrieve an overview rather than to aggregate specific evidence. Thus peer-reviewed research papers as well as  book chapters, 
were included and referred to as ``papers'' below.

We used IEEE Xplore, ACM, Scopus, Wiley, and Web of Science as the indexing services for finding the literature. Our search criteria “testing of autonomous systems” was applied to titles, abstracts and keywords. Titles and abstracts were examined for inclusion/exclusion based on relevance. 
We also used backward snowballing to track more relevant papers. In total, we identified 45 papers as listed in the complementary material~\cite{supplementary} in which the majority of them were acquired through the initial search results.


\subsubsection{Analysis and Collation}

The first author saved all selected papers into Zotero
, and studied the full-texts. For each of the papers, important segments discussing the features under study 
were highlighted, then short labels and, detailed notes if necessary, were created in Zotero. 
The labels were then refined and sorted to be coherent, consistent, and distinctive codes as unclear and duplicate entries were removed. Lastly, they were imported in XMind
for thematic synthesis based on the guidelines by Cruzes et al.~\cite{cruzes2011recommended}. A thematic model was created with 84 codes that organized around the three research questions 
and reviewed by the second and the third authors. Details of the thematic model can be found in the complementary material~\cite{supplementary}.

\subsection{Focus Group}

\subsubsection{Participants Selection}

We arranged a focus group \cite{daneva2015focus} discussion in April 2020 to get insights from industry practitioners. Participants were selected using convenient sampling \cite{ghazi2018survey} based on 
an invitation towards a network of testers in Southern Sweden. 
In total, 8 participants joined the focus group, as summarized 
in Table~\ref{table:partcipants_focus_group}.

\subsubsection{Implementation}

Due to the pandemic, the focus group was conducted via Zoom
. 
We started with a general introduction about this study and testing of autonomous systems; then the three research questions were discussed, each devoted about 30 minutes. For each question, first, participants discussed the question in breakout rooms with 2-3 persons each and wrote answers on Padlet
; second, all participants were brought back to the main session and the moderator led a discussion to elaborate and expand the answers on Padlet. The focus group was video recorded with consensus from all participants and the Padlet notes were saved by the end.

\begin{table}[!t]
\renewcommand{\arraystretch}{1.3}
\caption{Overview of participants in the focus group}
\label{table:partcipants_focus_group}
\centering
\footnotesize
\begin{tabularx}{\columnwidth}{@{}p{0.06\columnwidth}<{\centering}p{0.32\columnwidth}<{\raggedright}p{0.32\columnwidth}<{\raggedright}p{0.3\columnwidth}@{}}
\hline
\# & Position & Domain & Experience\\
\hline
1 & Manager & Health Care & 25-30 years\\
2 & Test Specialist & Software Engineering & 20-25 years\\
3 & Researcher \& Engineer & Software Engineering & 10-15 years\\
4 & Researcher \& Engineer & Logistics & 0-5 years\\
5 & PhD Cand. \& Engineer & Software Engineering & 0-5 years\\
6 & Professor & Artificial Intelligence & 20-25 years\\
7 & Senior Researcher & Software Engineering& 5-10 years\\
8 & PhD Candidate & Robotics & 0-5 years\\
\hline
\end{tabularx}
\end{table}

\subsubsection{Result Analysis and Synthesis}

We adopted the inductive thematic synthesis approach proposed by Cruzes et al. \cite{cruzes2011recommended} for coding. 
The first author conducted the primary coding. First, the recorded videos were reviewed, and notes were taken; Padlet answers were then combined to generate the codes. Second, the codes were imported and anlyzed in XMind for thematic synthesis. 
A thematic model was created with 37 codes that organized around the three research questions. 
The resulting model was reviewed by the second and the third authors, and can be found in the complementary material~\cite{supplementary}.

\subsection{Interviews}

\subsubsection{Participants Selection}

To validate and complement the findings from the previous activities, we conducted four interviews with industry practitioners that were not involved in the focus group. Now, we specifically approached experts in our network who had worked with autonomous systems. Five interviewees from industry accepted the invitation as shown in Table~\ref{table:interviews}, where \#3 and \#4 were interviewed together.
  
\subsubsection{Implementation} 

The interviews were conducted on Zoom by two of the authors each. 
Two of the interviews were 60 minutes in length and the other two lasted for 45 minutes. The interview schema was semi-structured, guided by Runeson et al. \cite{runeson2009guidelines}, and Rowley et al.~\cite{rowley2012conducting}. The interview questions, as listed in the complementary material~\cite{supplementary}, were derived from the literature review and focus group discussion. 
The interviews were video recorded with consent from the interviewees.

 
\subsubsection{Result Analysis and Synthesis} 

We used the same synthesis approach~\cite{cruzes2011recommended} as above. 
Recorded videos were first transcribed in Nvivo, 
 and important segments of the text were highlighted and coded, resulting in 62 codes. They were imported in XMind for thematic synthesis, where, duplicate and unclear entries were removed, and a thematic model with the codes was created and organized around the three research questions. 
 Lastly, the resulting model was reviewed by the second and third authors, and details of the model can be found in the complementary material~\cite{supplementary}. 
 
\begin{table}[!t]
\renewcommand{\arraystretch}{1.3}
\caption{Overview of participants of interviews}
\label{table:interviews}
\centering
\footnotesize
\begin{tabularx}{\columnwidth}{@{}p{0.06\columnwidth}<{\centering}p{0.35\columnwidth}<{\raggedright}p{0.25\columnwidth}<{\raggedright}p{0.34\columnwidth}@{}}
\hline
\# & Position & Domain & Experience\\
\hline
1 & Industrial PhD Candidate & Automotive & 5-10 years\\
2 & Solution Architect & Automotive & 5-10 years\\
3 & Lead Software Architect & Mobility & 20-25 years\\
4 & Software Architect & Mobility & 20-25 years\\
5 & Technical Manager & Manufacturing & 20-25 years\\
\hline
\end{tabularx}
\end{table}

\subsection{Final Thematic Model Synthesis}

The thematic models generated from the three activities were reviewed, refined, and further compared to eliminate any potential conflicts across these models in XMind. Then they were synthesized into a final thematic model. First, the distinctive themes and codes from the three models were identified and moved to the final model, then the rest of the themes and codes, which share the same or similar purpose, were analysed further to either be added to the final model or merged with other codes. 
Second, the final thematic model was reviewed to ensure that it integrates the themes and the codes from all three models, and resulting codes remain being coherent, consistent and distinctive. 

\subsection{Validity}
As reported above, we have used systematic research methods to improve the validity of the synthesized conceptual model. Being an exploratory study, aiming to understand concepts and practices, we value \emph{construct validity} highest.  
To strengthen the construct validity, we have asked open questions, not relying on predefined terms and concepts. 
The \emph{reliability} or \emph{trustworthiness} of the study is addressed through rigorous data collection and analysis procedures. The analysis of the data took place in three steps, each focusing on one source of empirical evidence (literature, focus group, and interviews) to ensure that the concepts may emerge from the empirical source, which thereafter were unified into one conceptual model. 
The \emph{external validity} is related to the scope of  the model. 
We have extended the scope beyond the most prevalent automotive domain, by searching the literature broadly, interviewing people from other domains, and explicitly asking for autonomy concepts in virtual only domains, like stock markets. However, we don't make any claims with respect to completeness of industry domains. 



\section{Results}
\label{sec:results}

The resulting conceptual model presents three different contributions of our study: 1) A conceptualization of autonomous systems, 2) A classification of challenges for testing of autonomous systems, and 3) A classification of available techniques and approaches as well as current practices for testing of autonomous systems. The following sub-sections explain the findings in a more detailed manner, including excerpts from our analysis model in Figures~\ref{fig_autonomous}--\ref{fig_techniques}.

\subsection{Conceptualization of Autonomous Systems}

\begin{figure}[!t]
\centering
\includegraphics[trim=0 25mm 0 27mm, clip, width=\columnwidth]{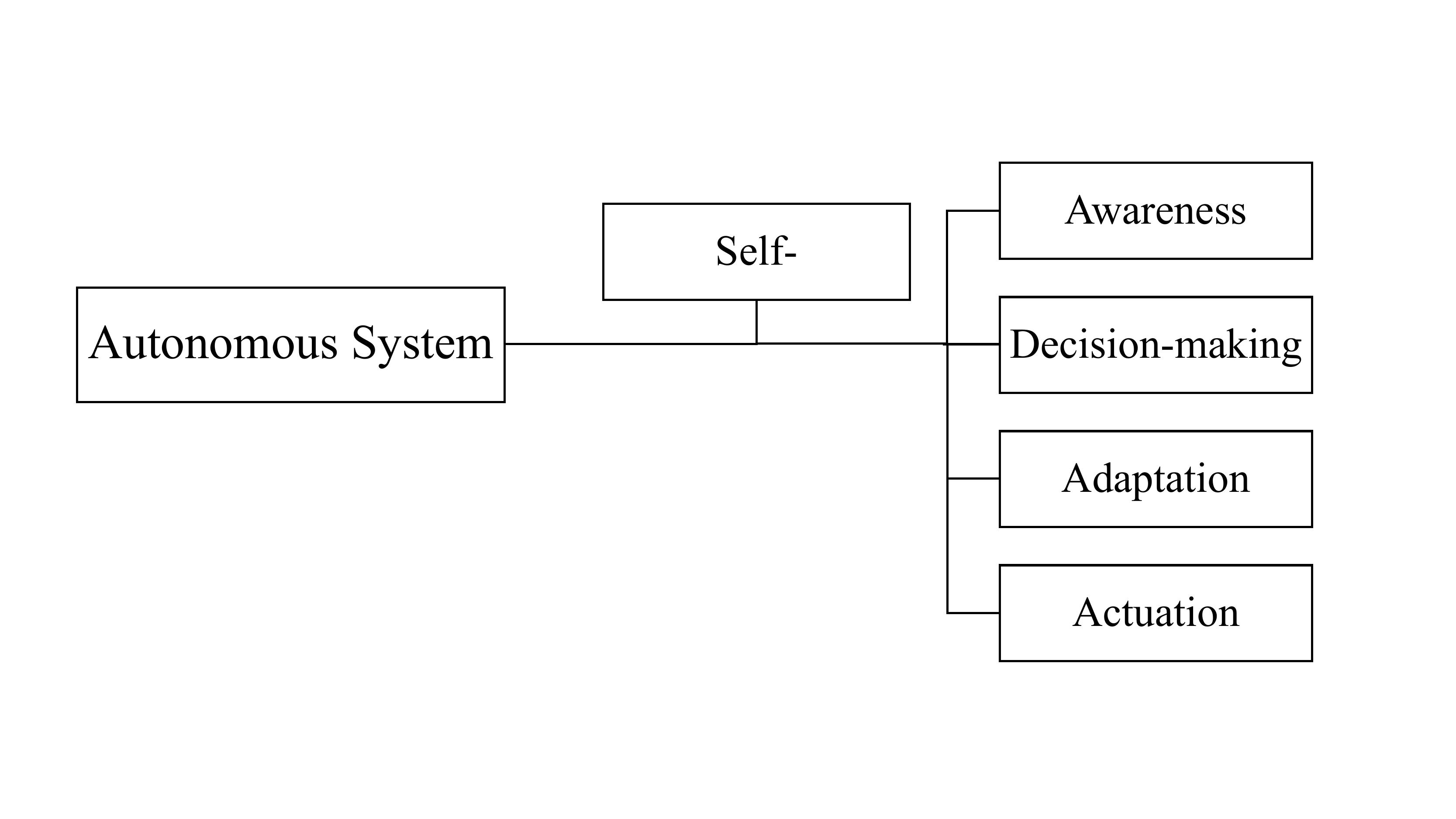}
\caption{Concepts of autonomous systems}
\label{fig_autonomous}
\end{figure}

As there is no universal definition of what constitutes an autonomous system, 
we synthesized the variety of notions used in literature and practice for different application domains. 
The result was a taxonomy of aspects defining an autonomous system, as presented in Figure~\ref{fig_autonomous}. Both the literature and industry practitioners similarly described that, autonomous systems \emph{are capable of performing certain tasks in an unstructured environment without human supervision}~\cite{helle2016testing}. More specifically, the system must be \emph{self-aware}, meaning that it can analyse the situation and do the \emph{decision-making} on its own, given the environmental conditions and its own states. Furthermore, the system must be able to \emph{actuate} the plans to fulfil the desired tasks, and \emph{adapt} its behaviour to optimize the goals by learning from the past.

A recent study by Sifakis \cite{sifakis2019autonomous} confirms our conceptualization by formalizing an architecture of autonomous systems, in which, the architecture defines five compulsory modules of an autonomous system: perception, reflection, planning, goal management, and self-adaptation. 

The mechanism of adaptation (rule-based or self-evolving) and appearance (physical or virtual) of the autonomous systems are two characteristics discussed
. While autonomous systems are expected to adapt their behaviour subject to conflicting goals and dynamically changing environments~\cite{helle2016testing, sifakis2019autonomous}, some interviewees emphasized the \emph{self-evolving capability}, without being explicitly implemented, as an essential nature of these systems. However, the others insisted on that rule-based systems can still hold some level of autonomy, in which, they are able to handle certain situations by themselves in a more limited way. As said by the interviewees \#3 and \#4:
\emph{We believe the rule-based systems can still be seen as autonomous systems as long as they offer the intelligence and autonomy in handling the tasks.}
The dispute on mechanism of adaptation becomes essentially a matter of the level of autonomy, as described in the automotive domain by the SAE 6-level of autonomous driving \cite{borg2018safely}. 

As for the appearance of the systems, the most intuitive cognition of autonomous systems involves the physical components such as sensors and electronics, as what has been integrated in vehicles and robotics. However, they can also appear in the digital form, e.g. smart software applications in the investment market \cite{sifakis2019can}, where they offer some intelligence, and react autonomously and virtually over non-physical media. This is a common view from the interviews and focus group discussion, as interviewee \#2 stressed: 
\emph{A pure software system can also be instance of autonomous systems, since it is still the software control units, which lie in the heart of the autonomous systems, that actually enable the system autonomy.}

Given the broad interpretation of autonomous systems in different industry contexts, ranging from automotive, robotics, aviation, healthcare, cyber-security, to smart software systems, a \emph{definition of autonomous system should incorporate systems both in a physical and digital form}. Besides, the definition should also be inclusive for different mechanisms of adaptation. A rule-based system can still generate some level of autonomy and handle unforeseen situations without human involvement, and self-evolution can be viewed as a feature that enables full autonomy where the system has the intelligence to reason, analyse and learn from both the surroundings and experiences, without being explicitly programmed during design.

\subsection{Challenges for Testing}

The challenges for testing of autonomous systems, as defined by our conceptualization model, come from two primary concerns, namely quality and safety. The quality aspects are committed to assuring the correctness of the design, the code, and the behaviour, while the safety aspects are about ensuring that potential incidents are within an acceptable threshold. Sifikas et al.~\cite{sifakis2019can} articulated that the machines must cope with the human order and should not expose any risk or danger to human society. Also, according to Helle et al. \cite{helle2016testing}: 
    \emph{Humans usually have high expectations for autonomous systems but low tolerance on their faults.}

\begin{figure}[!t]
\centering
\includegraphics[trim=0 25mm 0 27mm, clip, width=\columnwidth]{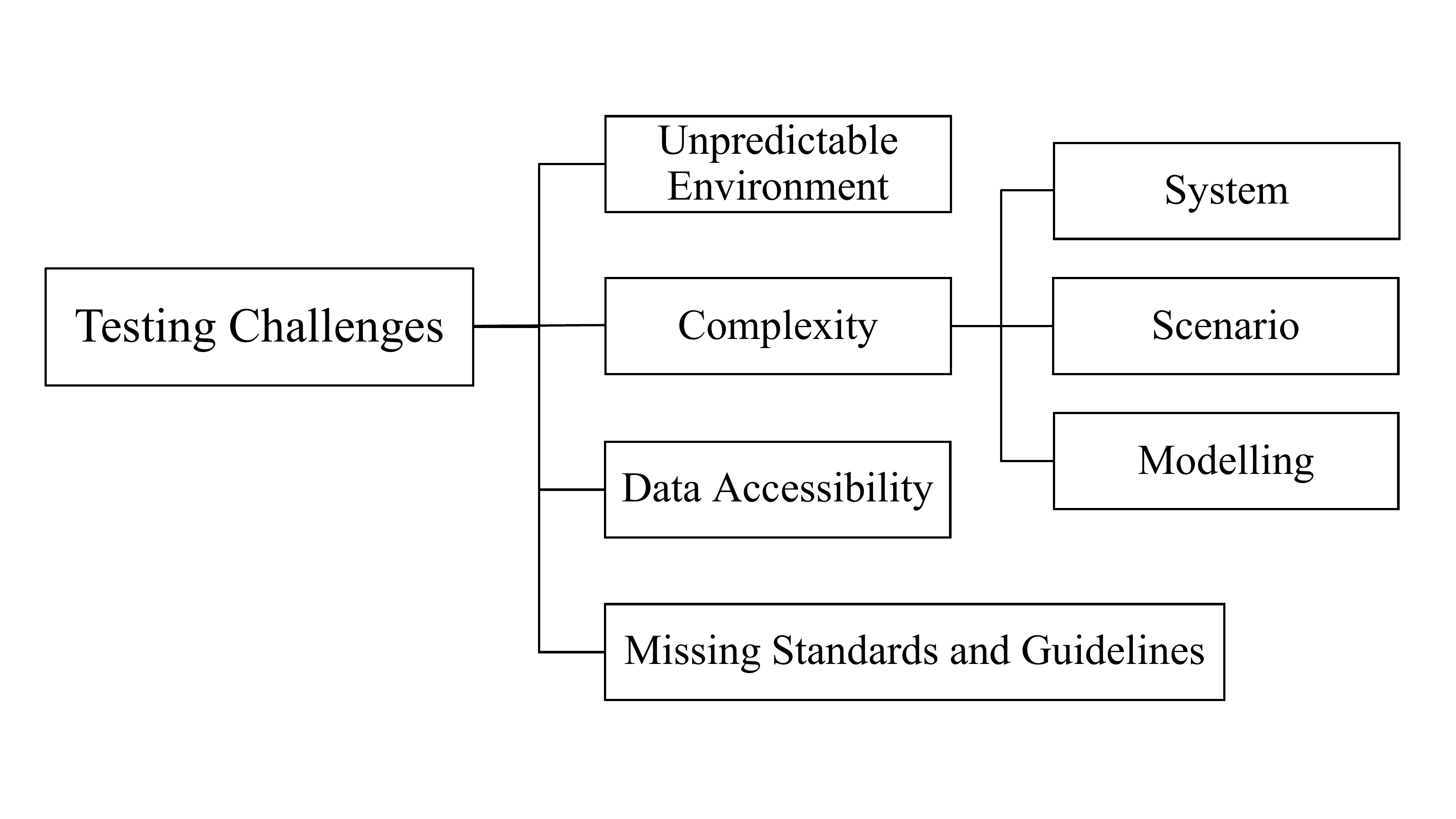}
\caption{Challenges for testing of autonomous systems}
\label{fig_challenges}
\end{figure}

 Unfortunately, our results indicate that the challenges on quality and safety of autonomous systems are far from being resolved due to the \emph{unpredictable environment}, the \emph{complexity}, \emph{data accessibility}, and \emph{no standards or guidelines} that are settled for testing, see Figure~\ref{fig_challenges}.

\subsubsection{Unpredictable Environment}

The unpredictable environment is one of the major impediments that add uncertainties to testing as the systems can run into any environmental conditions that were unknown during design. Further, the same input can lead to different results since the system will learn and adapt its behaviour after deployment~\cite{helle2016testing}. For example, a satellite has to respond to all risks in space for years and human intervention is impractical after launch. This leaves a large explosion of parameters to be explored and it is infeasible to cover all possible scenarios~\cite{helle2016testing, koopman2016challenges}. As reported in research on autonomous cars~\cite{koopman2016challenges, tao2019industrial, mazzega2016testing} and also argued by the interviewees, 
a number of million kilometers’ driving test for the vehicle would guarantee the safety based on statistic prediction and the level of ambition. However, in reality, it is too expensive to conduct that distance of driving test in the traffic within years and there might still be corner cases, low-frequency errors, that are not covered. 

\subsubsection{System and Scenario Complexity}

The complexity of testing autonomous systems lies in all artifacts involved in the operational environment and the system itself. 
The system is typically built as a system of systems, which involves many software control units and hardware electronic components. The emergence of AI technologies has increased the system complexity due to the limitations of existing techniques in addressing their test-ability, interpret-ability, and visualize-ability~\cite{borg2018safely}. 

The performance of these AI-enabled systems depends largely on the data, where the implementation provides only the pre-trained model and leaving the actual behaviour non-deterministic and subject to data acquired during operation. Therefore, conventional testing approaches such as unit testing, component testing, and code review are inadequate to ensure the quality and prevent, identify or remove undesired consequences~\cite{borg2018safely, paulweber2017validation, aniculaesei2018toward}. The testing has to ensure not only the correctness of the code and algorithms, but also the behaviour of the system that is determined by the actual input. This requires a large quantity of data, that are reliable and based on a real-world distribution, and a thorough understanding of how the systems are evolving. 

The black-box nature of deep learning algorithms makes it even harder to understand or visualize the process of the decision-making~\cite{borg2018safely, zhang2020testing} as it can go through hundreds of layers before reaching a certain decision, and millions of parameters can be involved in this process. Besides, the AI components are exposed to adversarial attacks that can mislead the data and the communications in between
~\cite{knauss2017paving}. A recent example is that a man with 99 mobile phones on a kid’s cart flawed Google Maps as a traffic congestion~\cite{google2020hack}.

\emph{Scenario} complexity is yet another challenge at the core of the testing of autonomous systems \cite{agaram2016validation}. It is hard to track, record, and replicate failures, particularly for fatal crashes or near-crashing cases~\cite{paulweber2017validation}. It is unclear how engineers can define a scenario that includes all artifacts, either in a real environment or a simulation environment~\cite{knauss2017paving, pfeffer2016continuous}. Thus a better understanding is needed, of which factors, or objects, in the testing scenario led to the existing consequence, and whether the scenarios used for testing reflect the actual situations on how human operators react. In addition, established terminology and tools are imminently demanded~\cite{agaram2016validation}.

As a result, the testing, which to a great extent is relying on the \emph{modelling} of the system, the environment, and the scenarios, is getting intricate due to the complexity of all of them. It leaves academia and industry to improve and invent tools and approaches on how to model the environment, the system, and scenarios to keep the simulation environment align with the real-world~\cite{knauss2017paving, paulweber2017validation}.

\subsubsection{Data Accessibility}

To be able to analyse, model, and test the systems, more data of good quality are required. 
However, the data becomes extremely costly when it comes to the collection, labelling, interpretation, validation, and generation of testing data~\cite{knauss2017paving, koopman2016challenges, bach2017data, tian2018deeptest}. The developers and testers must not only collect the data, but also understand the significance, dimensions, and distribution of the data in different formats, label and validate the data to not under-fit or over-fit the performance. On some occasions, the data engineers must generate reliable data to compensate lack of data for testing purposes. In addition, one of our interviewees also expressed that they were struggling with data ownership issues to acquire data access between organizations. 

\subsubsection{Missing Standards and Guidelines}

One problem that aggravates the complexity for testing of autonomous systems is that no standards and guidelines are settled
~\cite{helle2016testing, koopman2016challenges, paulweber2017validation}. 
Thus a new foundation must be established for autonomous systems~\cite{harel2020autonomics} to understand, e.g. how to specify the requirements with suitable terminology, what quality criteria and safety performance to adopt, what oracle to pass or fail the test cases, how to conduct the regression test if revisions are made during tests, and what policies, regulations and ethical standards to apply. 
New tools and approaches must be invented to guide and automate the testing in an efficient and effective way. Quotes from two interviewees state that 
    \emph{\#2: The industry is not prepared yet to address these issues into standards and set guidelines on how to do it.}
    \emph{\#5: More education and research are required to get the industry ready to mitigate the challenges and for the society to get along with the autonomous technologies and products.}

\subsection{Techniques, Approaches and Practices}

Existing techniques, approaches, and practices for testing autonomous systems are insufficient to address the testing in an efficient and effective way, whereas they still address some important testing perspectives. 
Most of our findings, as described in the following sub-sections, are extracted from the literature study since the industrial practitioners usually focused on one or two of the approaches.

\subsubsection{Practices -- Available Industry Standards}

Some industry standards are mentioned by the literature and industry practitioners from the focus group as well as the interviews, even though none of them specifies what is required for testing the fully autonomous systems. Among them, ISO-26262 addresses the functional safety for road vehicles~\cite{koopman2016challenges, wotawa2017testing}, which impacts on how automotive software is designed, developed, and tested. However, as more autonomous functionalities are involved and enabled by AI technologies, such as deep learning neural networks, the techniques within this standard such as code review and coverage-based testing are no longer applicable~\cite{borg2018safely}. 
The ISO-16787 specifies test procedures and performance requirements for assisted parking systems~\cite{meng2018parking}. It serves more as a suggestion and is up to each nation to implement. IEC-61508 introduces the fundamentals of functional safety for the electrical/electronic/programmable safety-related systems and focuses on the hazards caused by malfunctioning rather than any external environmental related factors~\cite{borg2018safely, zhang2020testing}. 
A newly published standard, which aims for the safety of the intended functionalities (SOTIF) for automotive, is described in ISO/PAS-21448 \cite{borg2018safely, zhang2020testing}. This standard provides guidance and measures needed for the applicable design, verification, and validation to achieve the SOTIF. 


\subsubsection{Practices -- Engineering Recommendations}

\begin{figure}[!t]
\centering
\includegraphics[trim=0 7mm 0 8mm, clip, width=\columnwidth]{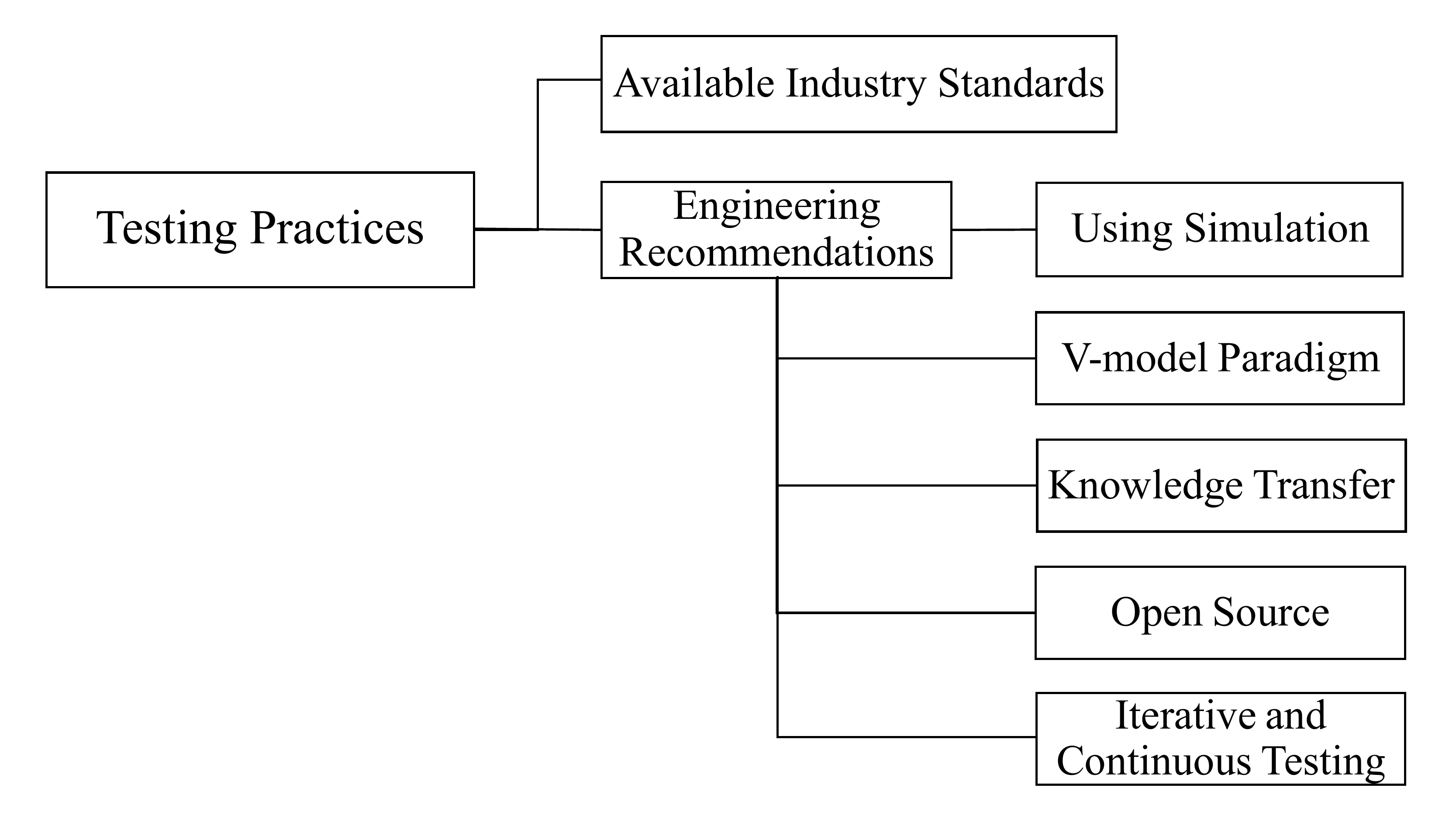}
\caption{Practices for Testing of Autonomous Systems}
\label{fig_practices}
\end{figure}

Several engineering recommendations, as shown in Figure~\ref{fig_practices}, were articulated both in the literature and by the industry practitioners, such as, \emph{using simulation} for testing of autonomous systems to reduce the cost of accessing expensive hardware facilities and the risk of generating safety issues and economic losses. At the same time, simulation can considerably improve the testing efficiency by monitoring the test, recording the data, and generating the failure or test report. Besides, simulation also benefits test analysis by visualizing the process of the test.

The \emph{V-model paradigm} is a common practice for test decomposition
~\cite{koopman2016challenges, huang2016autonomous}. In V-model engineering, 
software testing on different levels require different techniques and approaches, starting with unit testing, component testing, and integration testing, and then moving the entire system with both software and hardware parts into a simulation environment, to test ground and into the real-world to ensure that both the functional and non-functional requirements are satisfied. 

Our interviewees and focus group participants also emphasized that \emph{knowledge transfer} is essential. By learning from other industry domains and collaborating with both academia and industry, the entire industry should set up standards and regulations jointly, \emph{open source} the data and platforms for reusing instead of having every player creating its own. As said by the interview \#2:
    \emph{We must learn from the other industry domains and transfer knowledge across. We must also initiate the collaboration among the industry for reusing the data and the tools instead of creating your own.} 

Kang et al. presented 37 datasets and 22 virtual testing environments that are publicly available for closed-loop testing for autonomous vehicles~\cite{kang2019test}. Academic researchers can well contribute to explore possible alternatives, with the industry providing the test data, test-beds, and test results. Thus, industry and academia should move forward hand in hand. One of the most important and practical strategies for testing of autonomous systems is to go from requirement-driven engineering, 
to aim for \emph{iterative and continuous engineering}, where it may initially start with limited testing data and an incomplete testing model, as expressed by interviewee \#2: 
    \emph{We do not expect the testing can be solved with everything known beforehand, but rather taking it continuously in step-wise. The point is, when we start testing, we will get the data and we know better what is the problem.}

\subsubsection{Techniques and Approaches (Conventional)}

Conventional testing approaches are deemed as inadequate for addressing the autonomous nature of the systems~\cite{helle2016testing}, but they still dominate the testing efforts and resources for the time being. Regardless of the complexity of the system architecture, the software is still enabling the system intelligence and autonomy. The conventional software testing process, in brief words, can start with approaches such as unit testing, component testing, integration testing and non-functional testing. 
Depending on the testing and integration plans, the other components and sub-systems are then included and tested in simulation with Software-in-the-loop, Hardware-in-the-loop, and Vehicle-in-the-loop for the automotive~\cite{huang2016autonomous}. In the later stage of testing, it involves ground testing or test in production environment, such as, test track and real road testing for vehicles and mobile robots before deployment~\cite{agaram2016validation, huang2016autonomous}.

\subsubsection{Techniques and Approaches (Autonomy-focused)}

\begin{figure}[!t]
\centering
\includegraphics[trim=0 7mm 0 8mm, clip, width=\columnwidth]{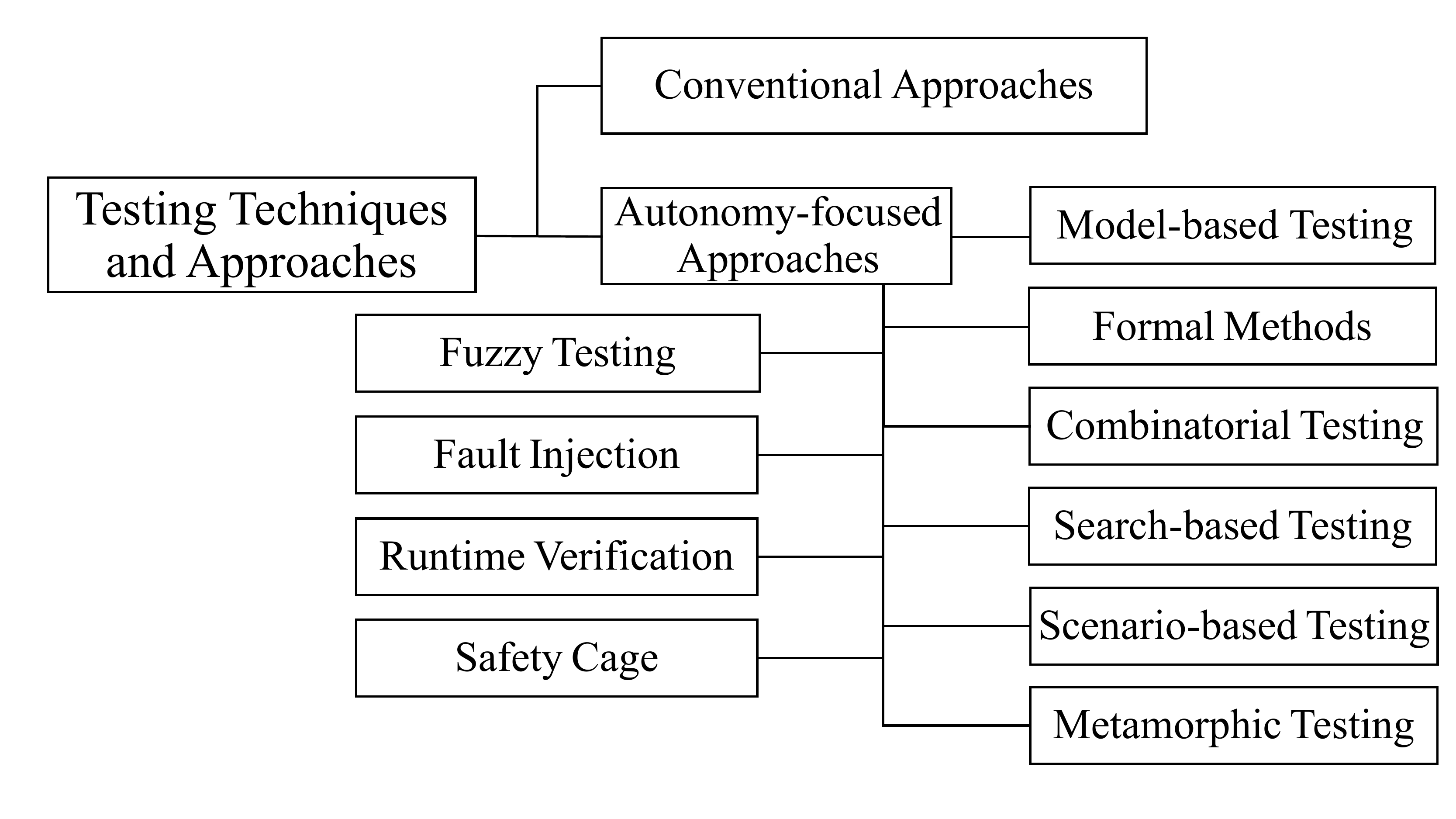}
\caption{Techniques and Approaches for Testing of Autonomous Systems}
\label{fig_techniques}
\end{figure}

There are testing techniques and approaches used for solving autonomy-derived issues, as proposed by existing academic research and industrial practices, and shown in Figure~\ref{fig_techniques}. \emph{Model-based testing} approaches~\cite{helle2016testing, aniculaesei2016towards} are used to model the system properties, constraints, and behaviour, and thus can be further utilized for automatic generation and execution of the test cases. \emph{Formal methods}~\cite{borg2018safely, huang2016autonomous, aniculaesei2016towards} is another similar way to analyse and represent the system design, inputs and outputs using domain terminologies, and validate the system against a formal specification. These techniques and approaches are commonly used for rule-based systems. 

Several other techniques are developed with promising results in reducing the number of test cases and total test efforts. Among these, \emph{combinatorial testing}~\cite{tao2019industrial, wotawa2017testing} combines and adjusts multiple parameters in one test scenario instead of having one parameter being updated with the rest remaining unchanged; \emph{search-based testing}~\cite{gambi2019asfault} and \emph{scenario-based testing}~\cite{Porres2020scenario} are used to explore and identify critical scenarios through statistical learning, e.g. using probabilistic models or genetic searching algorithms, where it analyses the previous testing results, studies the differences of them and approaches to the criticality objectives. 
Also, as highlighted by our interviewees \#2, that
    \emph{We think scenario-based testing is a very promising approach for testing of autonomous vehicles. Since it will be too expensive and impractical for us to cover all possible test cases, we should identify different scenarios instead, especially the worst-case scenarios, and put most of our test efforts in.}

\emph{Fuzzy testing}~\cite{zhang2020machine} and \emph{fault injection}~\cite{koopman2016challenges, hutchison2018robustness} are approaches that have been utilized to improve the test coverage and identify the corner cases, particularly for machine learning based applications. In detail, fuzzy-testing requires a large quantity of randomly selected data and validates the system performance based on the distribution and coverage of the test input. Fault injection is another variant where a set of special and faulty values are prepared to stimulate the systems and finding the corner cases. Another approach like DeepTest \cite{tian2018deeptest} was developed, where the researchers used image transformation to represent different real-world driving conditions and activate more neurons in the neural network as an indication for testing the autonomous driving algorithms. 

\emph{Runtime verification} include strategies, such as run time monitoring~\cite{mauritz2016assuring} and actuator-monitor architecture~\cite{aniculaesei2018toward}, which refers to that the system constantly monitors the behaviours during operation and report any anomalous situation as well as collecting data for reusing and optimization purposes. In addition, \emph{safety cage}~\cite{borg2018safely} is another approach that signals the anomalous inputs during operation by setting a confidence threshold and involving another control algorithm for situations below the threshold. 

In order to address the test oracle issue, \emph{metamorphic testing}~\cite{wotawa2017testing, lindvall2017metamorphic} was applied to define the metamorphic relations instead of specifying a certain value for asserting the test output. It is effective for many complex systems where the output of the test scenarios are hard to quantify but are consistent according to the inputs and certain principals, the metamorphic relations then act as the test oracle and expose a fault if the result fails to comply with them.

\section{Discussion}
\label{sec:discussion}


With the advent of autonomous systems, testing of such systems has become a challenge to practice as classical test approaches are not sufficient. Also the concept of \emph{autonomous systems} raises questions -- what do we mean by autonomy, and what kinds of systems may be labeled autonomous? Further, as this is an emerging field, with research and development spent in both industry and academia, it is of certain importance that the concepts are aligned to allow joint efforts and reduce the gap between industry and academia.


We therefore studied both academic literature and industry practice; the literature through a classical literature review, and industry practice through focus group discussions and interviews with practitioners. 
By synthesising a thematic model of the findings from different sources, we have presented an inclusive conceptualization of autonomous systems (RQ1).
We found a reasonable agreement on the \emph{autonomy} concept  to include aspects of performing tasks in unstructured environments without human supervision. However, whether or not the autonomy includes \emph{self-evolution} is not agreed upon. Further, while a lot of research and broader discussions on autonomous systems relate to physical systems, like robots and cars, both literature and practice confirm  that autonomous systems may be non-physical as well, for example, in banking and trade.

As a consequence, we propose that research be conducted on autonomous systems across domains, with physical as well as non-physical systems. Thereby general properties and techniques for autonomous systems may be developed rather than techniques for specific domains.



Given the characteristics of autonomous systems, we identified several \emph{challenges for testing} (RQ2). First and foremost, autonomous systems are expected to be able to meet \emph{unpredictable} situations and contexts, which by definition makes it impossible to test for a subset of such situations. Even when trying to specify example scenarios, they become very \emph{complex} or are not resembling reality. \emph{Access to data} is a key challenge, both to train and test autonomous systems, and since the field is emerging, \emph{standards and guidelines} for testing are not yet established. 

Implications for research and practice are that brute force traditional testing will never scale for autonomous systems. Rather, new approaches to modeling and simulation are needed, which align well with operational environments. Further, access to realistic data is a key for both training and testing. Most probably, domains have to collaborate on deriving and curating data.  

Even though the foundation for testing of autonomous systems seems weak, there are techniques and approaches used both in research and practice (RQ3). 
However, the academic contributions are mostly adaptations of approaches for conventional systems, and fewer novel approaches. Testing of autonomous systems do require novel approaches, but these may be well designed adaptations and combinations of elements already used for conventional systems.  

We do not claim that all kinds of autonomous systems will encounter the same challenges, nor that we have covered all kinds of systems. 
Nevertheless, we have explored the field of testing of autonomous systems and provided many insights based on both the academic publications and industry practices. Our results clearly indicate that the testing of autonomous systems is encountering a variety of challenges and must be improved aggressively. 

In the near future, we would like to extend the insights from more autonomous contexts, other than automotive, robotics, and manufacturing. We would also like to experiment with and compare the pros and cons of different techniques in real industrial contexts. As articulated by Harel et al.~\cite{harel2020autonomics}, a foundation for the next-generation autonomous system must be established, and according to Sifakis~\cite{sifakis2019can}:
    \emph{No power of decision to autonomous systems should ever be granted without rigorous and strictly grounded guarantees under the pressure of economic interests and on the grounds of ill-understood performance benefit.}

\section{Conclusion}
\label{sec:conclusion}

We have conducted an exploratory study on concepts related to testing of autonomous systems, through a semi-systematic literature review, a focus group and interviews with industry practitioners. We contribute to synthesizing 1) Concepts of autonomous systems, 2) Challenges for testing of autonomous systems, and  3) Techniques and approaches as well as practices for testing of autonomous systems. The findings are based on insights from both a literature review and industry practices, and can serve as a frame to facilitate, both academia and industry, on testing of autonomous systems. 

The conceptualization defines autonomous systems to be capable of performing specific tasks in an unstructured environment without human supervision. Our study also suggests that limited techniques and approaches have been reported so far by the academia and industry, and testing of autonomous systems must be substantially improved. The industry is trying different approaches without having common guidelines and standards, which is difficult and to advance the field, industry and academia must join forces in collaboration. 

\section*{Acknowledgment}

This work was partially supported by 
the Wallenberg AI, Autonomous Systems and Software Program (WASP) funded by the Knut and Alice Wallenberg Foundation.



%

\bibliographystyle{IEEEtran}
\bibliography{reference}

\clearpage




\end{document}